\documentclass[
    ,final            
  ]
  {aipproc}

\layoutstyle{8x11double}
\usepackage{hhline}
\usepackage{amssymb}

\begin{document}

\title{Onset of Turbulence in Superfluid $^3$He-B and its Dependence on Vortex Injection in Applied Flow}


\author{A.P. Finne}{
  address={Low Temperature Laboratory, Helsinki University of Technology,
P. O. Box 2200, FIN-02015 HUT, Finland} }

\author{R. Blaauwgeers}{
  address={Kamerlingh Onnes Laboratory, Leiden University,
P. O. Box 9504, 2300 RA Leiden, The Netherlands}}

\author{S. Boldarev}{
  address={Kapitza Institute for Physical Problems,
Kosygina 2, 119334 Moscow, Russia} }

\author{V.B. Eltsov}{
  address={Low Temperature Laboratory, Helsinki University of Technology,
P. O. Box 2200, FIN-02015 HUT, Finland}
  ,altaddress={Kapitza Institute for Physical Problems,
Kosygina 2, 119334 Moscow, Russia} 
}

\author{J. Kopu}{
  address={Low Temperature Laboratory, Helsinki University of Technology,
P. O. Box 2200, FIN-02015 HUT, Finland} }

\author{M.~Krusius}{
  address={Low Temperature Laboratory, Helsinki University of Technology,
P. O. Box 2200, FIN-02015 HUT, Finland} }


\begin{abstract}
  Vortex dynamics in $^3$He-B is divided by the temperature dependent
  damping into a high-temperature regime, where the number of vortices is
  conserved, and a low-temperature regime, where rapid vortex
  multiplication takes place in a turbulent burst. We investigate
  experimentally the hydrodynamic transition between these two regimes by injecting
  seed vortex loops into vortex-free rotating flow. The onset temperature of
  turbulence is dominated by the roughly exponential temperature dependence of vortex
  friction, but its exact value is found to depend on the injection method.
\end{abstract}

\def\XFMauthorsandtext{\\and~}
\maketitle


\emph{Introduction.} Superfluid $^3$He-B is the best medium to
study the influence of friction in vortex dynamics. The friction
arises when the superfluid vortex moves with respect to the flow
of the normal component. It consists of the longitudinal
dissipative and the transverse reactive contributions,
characterized by the mutual friction parameters $\alpha$ and
$\alpha'$ in the equation for the vortex line velocity
$\mathbf{v}_{\rm L} = \mathbf{v}_{\rm s} + \alpha
\hat{\mathbf{s}}\times (\mathbf{v}_{\rm n} - \mathbf{v}_{\rm s}) -
\alpha' \hat{\mathbf{s}} \times [\hat{\mathbf{s}} \times
(\mathbf{v}_{\rm n}-\mathbf{v}_{\rm s})]$.  Here
$\hat{\mathbf{s}}$ is a unit vector parallel to the vortex line
element. The velocities of the normal and superfluid fractions are
$\mathbf{v}_{\rm n}$ and $\mathbf{v}_{\rm s}$, while the
difference $\mathbf{v} = \mathbf{v}_{\rm n} - \mathbf{v}_{\rm s}$
is called the counterflow velocity, the hydrodynamic drive. An
important feature of $^3$He-B hydrodynamics is the large viscosity
of the normal component. Transient processes in the normal flow
decay quickly and can be neglected. Owing to this simplification,
with some modification the results for rotating flow can be
carried over to other types of flow.  With careful design and
preparation of the sample container the energy barriers preventing
vortex formation can be maintained high, so that high-velocity
vortex-free flow can be achieved.

The dynamics of quantized vortex lines can be explored if one
injects vortex loops in this meta-stable high-energy state of
vortex-free flow. The fate of the seed loops depends on
temperature: At high temperatures the number of vortices $(N)$ is
conserved, the injected loops expand to rectilinear lines, and the
flow relaxes only partially. At low temperatures below some onset
temperature, rapid vortex proliferation from the seed loops takes
place in a transient turbulent burst which leads to the formation
of the equilibrium number of vortices $(N_{\rm eq} \sim 10^3)$ and
to a complete removal of the applied flow \cite{nature}. At
sufficiently low temperatures $(T \lesssim 0.45 \, T_{\rm c})$ the
turbulent burst always follows, even after the injection of a
single vortex ring, independently of the injection method.
However, in a narrow temperature regime around the onset of
turbulence the situation is different: the injection may or may
not lead to turbulence depending on the injection details and the
velocity of the applied flow.

In this report we examine the dependence of the onset temperature
on the injection properties. The motivation is the following: A
number of different processes with their individual energy
barriers have to be traversed sequentially before turbulence in
the bulk volume becomes possible. The first is vortex nucleation.
It is here avoided by the injection of the seed loops. Next
follows a series of events which build up the vortex density
locally somewhere in the bulk volume for turbulence to start.
These processes act at the container wall. They have been explored
in Ref.~\cite{vortmult}. Compared to turbulence in viscous fluids,
the path leading to turbulence in superfluids appears to be more
straightforward to reconstruct.

\emph{Injection methods.} In our experiment \cite{turbJLTP} the
flow is created by rotating a cylindrical sample of 110\,mm
length and $R = 3\,$mm radius around its symmetry axis. We employ
four different techniques to inject vortex loops in the rotating
flow, in order to study the transient evolution from the
vortex-free state to a final stable state with a central vortex
cluster consisting of rectilinear vortex lines. The final state is
only meta-stable, unless the cluster contains the equilibrium
number $N_{\rm eq}$ of vortex lines. With NMR techniques we
measure the number of lines $N$ close to both ends of the sample.
At temperatures above the onset of turbulence one observes regular
expansion of the injected loops to rectilinear lines, i.e. in the
final state $N \ll N_{\rm eq}$. At temperatures below onset the
injection evolves into a turbulent burst, which results in a
large increase in $N$, so that in the final state $N
\approx N_{\rm eq}$.

\begin{table}
\caption{Characterization of vortex injection methods in rotating
superfluid $^3$He-B.} \small
\begin{tabular}{l|c|c|c|c}
\hhline{-|--|--|} \hfil\textbf{Method} & \textbf{Trigger} &
\textbf{Number of loops} & \textbf{Length scale} &
\textbf{Location in sample container}\\
\hhline{-|--|--|}
Kelvin-Helmholtz  & Slow $\Omega$ sweep            & smooth distribution               & 0.1 -- 1 mm,          & At AB interface close to \\
instability of AB & across $\Omega_{\rm c}(T,P,H)$ & $N\sim 3$\,--\,30 (peak $\sim$ & a wavelength & cylindrical wall \\
interface \cite{turbJLTP}         & $\sim 0.8$\,--\,1.6 rad/s      & 8,
long tail up to 30)            & of ripplon               & at $R_{\rm AB}\approx 2.6\,\mbox{mm} < R$ \\
\hhline{-|--|--|}
Nucleation of $^3$He-A & Slow sweep of & & Circumference of & At newly forming AB \\
in magnetic field & barrier field up to & $1 \ll N < N_{\rm eq}$ & AB interface along & interface \\
from $^3$He-B & $H_{\rm AB}(T,P)$ & & cylindrical wall \\
\hhline{-|--|--|}
Neutron absorption & Neutron absorp- & $N \sim 1\,$--\,5 & $\sim 100\,\mu$m (diameter & Random location close to \\
\cite{pltp,neutronJLTP} & tion event & depends on $\Omega / \Omega_{\rm cn}$ & of largest vortex ring) & cylindrical wall\\
\hhline{-|--|--|}
Remnant vortex \cite{vortmult} & Rapid increase of & $N \sim 1$ & $\sim  R$, size of remnant & Random and distributed \\
 & $\Omega$ from zero & & vortex loop at $\Omega = 0$ & along sample \\
\hhline{-|--|--|}
\end{tabular}
\label{injtab}
\end{table}

\begin{figure}
\includegraphics[width=\linewidth]{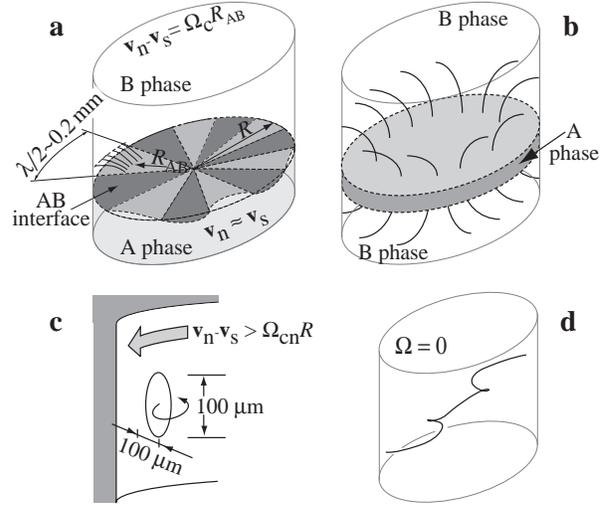}
\caption{Vortex loop configurations at injection in a rotating
cylinder: {\bf (a)} Two-phase
  sample with roughly the equilibrium number of vortices in A
  phase and vortex-free B phase. The A phase vortices curve at the AB
  boundary on the interface, forming there a surface layer of vorticity. In
  the KH instability the A-phase vortices in the deepest trough of the
  interface wave are tossed on the B-phase side
  in the region between $R_{\rm AB} \approx 2.6\,$mm and the cylinder wall
  at $R = 3\,$mm. {\bf (b)} When the barrier field is slowly swept up (at
  constant $\Omega$, $T\!$, $P$) the nucleating A-phase layer is unstable
  and a massive number of vortex loops escapes to the B phase. {\bf (c)} In
  a neutron absorption event close to the cylinder wall vortex rings are
  extracted by the applied flow from the reaction heated volume. {\bf (d)} A
  remnant vortex, which has not yet had sufficient time to annihilate at $\Omega = 0$,
  forms a seed for new vortex formation
  when the applied flow velocity ${\rm v} = \Omega R$ is increased from
  zero. }
\label{injfig}
\end{figure}

The different injection techniques are compared in
Fig.~\ref{injfig} and  Table~\ref{injtab}. The number of injected
vortex loops, their size, proximity, and the initial vortex
density at the injection site vary from one injection technique to
the next. We would like to answer the question whether these
properties, besides temperature and drive velocity ${\rm v} =
\Omega R$, influence the onset to turbulence.

The first two injection methods are the most reproducible. They
are based on the properties of the first order interface between
the A and B phases of superfluid $^3$He. The A phase is stabilized
with a magnetic barrier field over a short section in the middle
of the long sample. The AB interface undergoes an instability of
the Kelvin-Helmholtz type when flow is applied parallel to it. As
a result of the instability vortex loops are tossed across the
interface from the equilibrium vortex state in A phase to the
vortex-free B phase.  The KH instability can be triggered (1) by
sweeping the rotation velocity $\Omega$ up to a well-defined
critical threshold $\Omega_{\rm c}(T,P)$ or (2) by sweeping at
constant $\Omega > \Omega_{\rm c}$ the magnetic field $H$ from a
low value up to where the A phase is suddenly nucleated with some
magnetic hysteresis at $H > H_{\rm AB}(T,P)$.

The third injection method makes use of rapid localized heating in a
neutron absorption event. From the overheated volume, which is $\sim
50\,\mu$m in diameter, one or more vortex rings may evolve above a critical
threshold $\Omega_{\rm cn}(T,P)$. The number of rings depends on the
applied flow ${\rm v} = \Omega R$ \cite{pltp}:
Just above $\Omega_{\rm cn}$ only one vortex loop is created per
absorption event, but with increasing $\Omega$ the average number of loops
per injection event increases, reaching $\langle N \rangle \approx 5$ at
$\Omega/\Omega_{\rm cn} = 4$.

In the fourth method we start from an existing remnant vortex and
create the flow later. In a strict sense this is not injection as
in the three earlier cases, but in practice it achieves the same
result of placing a curved vortex loop in applied flow. With
decreasing temperature the last one or two vortices require an
ever longer time to annihilate at $\Omega = 0$ because of the
rapidly reducing vortex damping. Thus it becomes possible to catch
a remnant vortex before its annihilation in a random location at
the cylindrical container wall. When $\Omega$ is suddenly
increased to some final stable value, we may monitor the stability
of the remnant vortex in the applied flow \cite{vortmult}.

The variations in the injection properties prove to be larger
between the different methods than the variability within one
method from one run to the next.  We therefore list the
perturbations, which the different injection techniques generate,
in an order from the strongest to the weakest: {\bf (1)} First
comes the nucleation of $^3$He-A during a magnetic field sweep,
where the injected loop number is typically in the few hundreds.
Next comes {\bf (2)} the KH instability of the AB interface, {\bf
(3)} followed by neutron absorption at higher flow velocities.
{\bf (4)} Neutron absorption at lower velocities $\Omega \gtrsim
\Omega_{\rm cn} \sim 1\,$rad/s generates only one ring. The
weakest perturbation is {\bf (5)} a remnant vortex which can be
studied down to low flow velocities. The absolute limit is the
velocity at which a loop with a radius of curvature $\lesssim R$
is able to expand: ${\rm v} \gtrsim \kappa/(4\pi R) \,
\ln{(R/\xi)}$ which corresponds to $\Omega \sim 10^{-2}\,$rad/s.
(Here $\kappa$ is the circulation quantum and $\xi$ the coherence
length.) In practice the limit is set by our NMR detection which
at present requires $\Omega \gtrsim 0.4\,$rad/s \cite{turbJLTP}.

\emph{Onset of turbulence.} Let us examine with our different
injection methods the probability of turbulence within the
transition regime. We first consider the situation at a higher
temperature of $0.53 \, T_{\rm c}$: {\bf (1)} Here the nucleation
of $^3$He-A with magnetic field always gives the equilibrium
vortex state. {\bf (2)} Also KH injection has a high probability
$p_{\rm AB} = 0.96$ to start turbulence ($P=29\,$bar, $\Omega =
0.8$--1.6\,rad/s) \cite{turbdiagr}. In contrast, {\bf (3)} for
neutron absorption at $\Omega = 2.32\,\Omega_{\rm cn}$ the
probability is $p_{\rm n} = 0.09$ (Fig.~\ref{nprobab}). {\bf (4)}
No turbulent bursts have been observed with neutron injection, if
$\Omega < 2 \Omega_{\rm cn}$. {\bf (5)} Similarly vortex
multiplication is not triggered by a remnant vortex at this
temperature (at $P=29$ or 10 bar). The different injection
processes thus yield different onset temperatures for turbulence.

When temperature and mutual friction damping decrease, the magnitude of the
flow perturbation, which is needed to start turbulence, decreases as well. At
$0.45\, T_{\rm c}$ $(P=29\,$bar), vortex injection via {\bf (1)} the
nucleation of the A phase or {\bf (2)} the KH instability always result in
a turbulent burst. {\bf (3)} Similarly neutron absorption at high flow
$\Omega > 2 \Omega_{\rm cn}$ also has unit probability to initiate
turbulence. {\bf (4)} Even at lower flow velocity $\Omega \gtrsim
\Omega_{\rm cn}$, where only a single vortex ring can be injected from the
neutron bubble, the probability of obtaining turbulence is $p_{\rm n} =$
0.9 -- 1. {\bf (5)} A remnant vortex gives a probability of about 0.9
($P=10\,$bar) \cite{vortmult}. We concede that at low temperatures even the
smallest perturbation, a single quantized vortex loop, will evolve to a
turbulent tangle.

\begin{figure}
\includegraphics[width=0.9\linewidth]{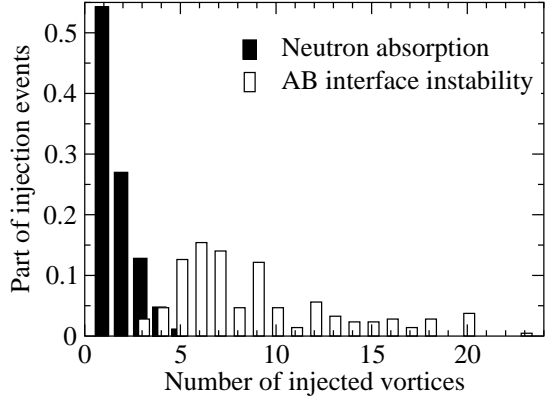}
\caption{Two histograms of the number of events (vertical scale)
in which a specified number of vortex loops (horizontal scale) is
injected in rotating vortex-free B-phase flow. These histograms
have been measured above the onset temperature of turbulence in
the regime of regular vortex dynamics. The example of KH injection
from the AB interface has been measured at $P=29\,$bar,
$T=0.77\,T_{\rm c}$, and $\Omega = 1.34\,$rad/s. Our measurements
indicate that this distribution is not strongly temperature or
velocity dependent. The example for injection via neutron
absorption represents an interpolation for $\Omega =
2.32\,\Omega_{\rm cn}$ from data measured at $T\approx 0.95 T_{\rm
c}$ and $P=2$--18\,bar (from Ref.~\cite{pltp}).  The number of
loops in neutron injection is strongly velocity-dependent, but
only weakly $T$ and $P$ dependent. } \label{distrib}
\end{figure}

\begin{figure}
\includegraphics[width=0.9\linewidth]{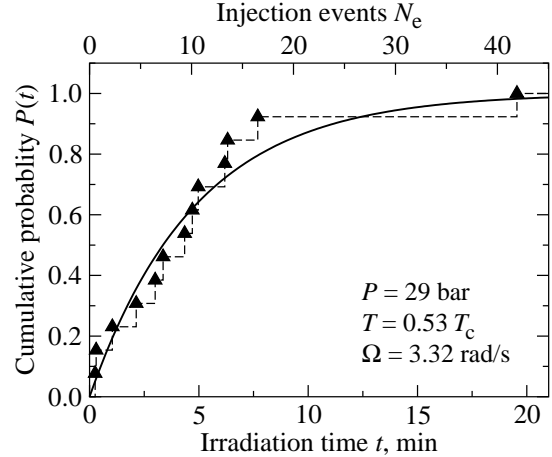}
\caption{Probability of turbulence after vortex loop injection from neutron
  absorption events at $\Omega = 2.32\,\Omega_{\rm cn}$. Each triangle
  represents a continuous neutron irradiation session of an originally
  vortex-free sample which, after a number of non-turbulent vortex
  injection events, terminates in a turbulent burst at the marked time. The
  triangles are distributed equidistantly along the vertical axis between 0
  and 1 to form approximation (dashed line) for the probability $P(t)$ to
  observe turbulence within the time span $t$ after starting irradiation.
  If each injection event has a probability $p_{\rm n}$ to generate
  turbulence independently, then $P(t) = 1 - (1 - p_{\rm n})^{\dot N_{\rm
      e} t}$. Here the average rate of injection events $\dot N_{\rm e}
  = 2.14\,\mbox{min}^{-1}$ is determined from the measured vortex
  formation rate \cite{neutronJLTP} and the average number of vortices
  formed in a neutron absorption event in Fig.~\ref{distrib}. A fit of the
  measured data to the expected dependence $P(t)$ gives the fitted
  probability distribution (solid curve) with $p_{\rm n} = 0.092$.  }
\label{nprobab}
\end{figure}

As seen from Table~\ref{injtab}, the various injection methods
differ in a variety of ways. The intensity of the perturbation,
which they present to vortex-free flow, arises from a combination
of different properties. A most important characteristic is the
number of injected loops. This is illustrated as a probability
distribution in Fig.~\ref{distrib} for two particular cases of
injection, namely via KH instability and neutron absorption.
Clearly the above examples from injection experiments at 0.53 and
$0.45\, T_{\rm c}$ support the simple notion that the more vortex
loops are initially injected, the larger is the probability of
turbulence. To explain these observations, we have to assume that
at $0.53 \, T_{\rm c}$ one needs to inject 4 -- 5 loops to achieve
turbulence, while at $0.45\, T_{\rm c}$ a single loop suffices. A
characterization of the injection methods in Table~\ref{injtab} in
terms of the number of injected loops would seem like a gross
oversimplification which ignores other differences, like the size
of the volume where the initial perturbation is localized. Still,
in the light of our results it looks conceivable that the number
of injected loops is a reasonable first measure of the intensity
of the flow perturbation.

\emph{Discussion.} With each injection method the transition from
regular vortex dynamics at high temperatures to turbulent vortex
dynamics at low temperatures occurs in a temperature interval of
certain width. We attribute this width to the variability in the
injection from one time to the next, since each injection method
is characterized by a certain distribution of configurations of
injected loops. Only some of these configurations result in
turbulence at higher temperatures. With decreasing temperature
more configurations become effective and eventually at low enough
temperatures all configurations produced by a given injection
method result in turbulence. The typical full width for a given
injection method is $\sim 0.06 \, T_{\rm c}$
\cite{turbdiagr,vortmult}. This width is generally smaller than
the difference between the average onset temperatures measured
with different injection methods.

Stability considerations of steady-state turbulence lead to the
conclusion that on average the transition between regular and
turbulent vortex motion is given by a condition on the ratio of
the mutual friction coefficients: $q = \alpha / (1-\alpha') \sim
1$ \cite{nature,transtheory}. Our measurements with different
injection methods and at different external conditions of $T\!$,
$P$, $\Omega$ show that the onset temperature is approximately
predicted by this criterion, but the exact value of temperature
does not correspond to any universal critical value of $q$.
Whether $q$ may still have a universal critical value in the
particular case of sustained homogeneous well-developed turbulence
is unclear, since no such measurements exist in the relevant range
$q\sim 1$.

The dependence of the onset on the intensity of the flow
perturbation resembles that observed in recent measurements
\cite{mullin} on the flow of a classical viscous liquid in a
circular pipe. Here a perturbation of finite magnitude $\epsilon$
needs to be injected in the flow, to turn it from laminar to
turbulent. A scaling law connects the smallest possible
perturbation $\epsilon$ and the Reynolds number: $\epsilon \propto
{\rm Re}^{-1}$, i.e. the minimum perturbation decreases with
increasing flow velocity. In superfluids  the analog of the
Reynolds number Re is $q^{-1} = (1-\alpha')/\alpha$
\cite{nature,turbJLTP}. Its magnitude increases monotonicly with
decreasing temperature (but does not depend on flow velocity).

The hydrodynamic transition from laminar (regular) to turbulent
flow in viscous liquids and in superfluids differs from a usual
first order phase transition, such as {\emph eg.} the transition
from supercooled meta-stable A phase to the stable equilibrium B
state of superfluid $^3$He. A single energy barrier, the creation
of a sufficiently large seed bubble of B phase with a critical
radius of order $\sim 1\,\mu$m, prevents the A$\rightarrow$B phase
transition. If such a seed bubble is injected, then the stable B
phase is irreversibly created in all of the available volume. If
the seed is too small, then it will shrink and disappear. In
viscous liquid flow along a circular pipe \cite{mullin}, an
injected perturbation creates turbulent ``puffs'' and ``slugs'' in
the laminar flow, which are limited in space and do not extend
over the whole length of flow. In our superfluid experiments an
analogous case appears if vortex multiplication in a turbulent
burst stops before the vortex number has reached $N_{\rm eq}$ and
large applied flow remains. Such intermediate events are rare, but
have been observed in the onset regime: In the middle of the
transition region their proportion among the well-characterized
events ({\emph i.e.} those which conserve the number of injected
vortices and those where the meta-stable flow is completely
removed) is at most a few percent. The existence of such
intermediate behavior is a further demonstration that in
hydrodynamic transitions often there is no single well-defined
energy barrier which one should overcome to cause a spontaneous
change from a meta-stable to a stable flow pattern
\cite{vortmult}.

\emph{Conclusion.} The onset temperature of turbulence after the
injection of vortex loops in meta-stable vortex-free flow depends
on several variables, but the decisive factor appears to be the
number of injected loops. When the number of loops increases they
are more likely to become unstable towards turbulence. As a result
the onset temperature of turbulence increases and the value of the
friction parameter $q^{-1}$ at onset decreases. This connection
between the amplitude of the flow perturbation and the ``Reynolds
number'' $q^{-1}$ at onset resembles the scaling law of classical
turbulence \cite{mullin}.


\begin{thebibliography}{9}

\bibitem{nature} A.P. Finne \emph{et al.},
\emph{Nature} \textbf{424}, 1022 (2003).

\bibitem{vortmult}  A.P. Finne \emph{et al.},
preprint \emph{arXiv:cond-mat/0502119}.

\bibitem{turbJLTP} A.P. Finne \emph{et al.},
\emph{J. Low Temp. Phys.} \textbf{136}, 249 (2004).

\bibitem{pltp} V.B. Eltsov, M. Krusius, and G.E. Volovik, in
\emph{Prog. Low Temp. Phys.} Vol. XV (Elsevier, Amsterdam, 2005).

\bibitem{turbdiagr} A.P. Finne \emph{et al.},
\emph{J. Low Temp. Phys.} \textbf{138}, 567 (2004).

\bibitem{transtheory} N.B. Kopnin, \emph{Phys. Rev. Lett.} {\bf 92}, 135301
  (2004); V.S. L'vov \emph{et al.}, \emph{JETP Lett.} {\bf 80}, 479 (2004);
  W.F. Vinen, \emph{Phys. Rev. B} {\bf 71}, 024513 (2005).

  \bibitem{neutronJLTP}  A.P. Finne \emph{et al.},
\emph{J. Low Temp. Phys.} \textbf{135}, 479 (2004).

 \bibitem{mullin} B. Hof \emph{et al.},
\emph{Phys. Rev. Lett.} \textbf{91}, 244502 (2003).


\end{thebibliography}
\end{document}